\documentclass{aa}  

\usepackage{graphicx}
\usepackage{txfonts}
\usepackage{url}
\usepackage{subcaption}
\usepackage{lscape}
\usepackage{placeins}
\usepackage{amsmath}
\usepackage{color}

\begin{document}

   \title{Distinct gas and stellar rotation curves in the Milky Way Galaxy}

   \author{A.\,H. Nelson\inst{1}\fnmsep\thanks{Corresponding author: e-mail: nelsona@cardiff.ac.uk}
   \and Y. Sofue\inst{2}
   \and P.\,R. Williams\inst{1}
   }

   \institute{School of Physics \& Astronomy, Cardiff University, CF24 3AA, UK
   \and
   Institute of Astronomy, The University of Tokyo, 2-21-1 Mitaka, Tokyo 181-0015, Japan
   }

   \date{Received ; accepted }

  \abstract
   {The rotational velocity of interstellar gas in the Milky Way, and other galaxies, has been taken to represent the circular velocity of a test particle in the Galaxy's gravitational field, and hence an indicator of the Galaxy mass. The derived circular velocity is found to be too high for the gas to be gravitationally bound to the galaxy given the observed Galaxy mass in stars and gas, and consequently an extra component of mass in the Milky Way and other galaxies, namely dark matter, has been postulated. However recently the observational satellite Gaia has been carrying out ground-breaking astrometric observations to accurately measure, inter alia, the three dimensional velocities of stars in the vicinity of the Sun and beyond. This has revealed that the circular velocity derived from the stellar population is much less than that of the gas, and the rotation curve, circular velocity versus radius, is distinctly declining with radius, whereas the gas rotation curve is not declining.}
   {This difference has previously been ascribed to inaccuracies and unreliability of the Milky Way gas rotation velocity measurements, however here we show that in fact the difference in the derived circular velocities is real.}
   {By combining results from multiple observations of the gas velocity, averaging the velocities in radial bins, we establish that there is a grand average rotation curve. This can be compared directly with a grand average of the published Gaia rotation curves, and the confidence level in the difference between the two estimated by statistical analysis.}
   {The difference is shown to have a high degree of confidence, and increases with galactocentric radius.}
   {The lower circular rotation curve from the stellar velocities has resulted in significantly reduced estimates of the dark matter mass fraction of the Milky Way. The higher rotation of the gas lacks an explanation, but it is unlikely to be an accurate indicator of the kinematic mass of the Galaxy. This also has significant consequences for the mass of external galaxies based on gas rotation curves.}


   \maketitle
   \nolinenumbers

\section{Introduction}
\label{sec:intro}

\subsection{Gas observations}
\label{sec:gas_obs}

Since the 1970s and 1980s the rotation curves of external galaxies and the Milky Way Galaxy have been obtained by observations of the interstellar gas, either directly via HI and CO emission, or indirectly via the H$\alpha$ line from HII regions around young stars, recently formed from the interstellar gas \citep{rubin1978,sofue2001}. Recently maser emission has also been detected from interstellar gas \citep{reid2019,hirota2020}. Many of these observations show a near universal flat rotation curve for the gas out to several times the effective radius of the galaxy. The local turbulent fluctuations of gas velocity in the Milky Way are relatively small compared to the rotation velocity, although systematic modulations due to shock and streaming motions in spiral arms exist. Consequently, where streaming motions can be accounted for, the gas velocity is deemed to be close to the circular velocity of a test particle in the Galaxy gravitational field. For the Milky Way, measurement of the rotational velocity beyond the solar radius is difficult since we are situated internal to the gas disc, however the flat gas rotation curves with velocity ${\sim}\,250$~km\,s$^{-1}$ have been confirmed out to a radius ${\sim}\,25$~kpc, and for external galaxies the flat gas velocity rotation curve has been detected far beyond this radius.

On the assumption that the gas is bound to the Galaxy by gravitation, such high velocities would require more mass than has been detected in the Galaxy. Including the stellar disc and bulge and the interstellar gas the detected mass inside ${\sim}\,20$~kpc is estimated to be $(0.6$--$1.0)\times 10^{11}\,M_\odot$ \citep{nicastro2016}, but to contain the gas velocities a total mass of ${\sim}\,2 \times 10^{12}\,M_\odot$ would be required \citep{wang2023,moffat2024}. Consequently an unseen component in the form of dark matter has been postulated to exist with a mass of ${\sim}\,2 \times 10^{12}\,M_\odot$. Similar massive dark matter components have been postulated for external galaxies on the basis of their gas rotation curves.

\subsection{Gaia observations}
\label{sec:gaia_obs}

Until recently estimating the circular velocity associated with the older stellar disc of the Galaxy has been difficult since, unlike the gas component which has relatively small velocity dispersion, the older stellar component has a significant velocity dispersion. Measurement of the three dimensional velocities of the stars is needed to calculate the corresponding circular velocity of the stellar disc. Gaia was designed to measure the 3-d velocities of nearby stars, among other quantities, by combining measured proper motions with radial motion with respect to the solar system. In 2023 the Third Data Release (DR3) of over 33 million stars was made public, and since then the velocity data of Population~II stars, i.e.\ older stars with no recent direct connection to the interstellar gas, have been analysed to obtain the circular velocity derived from this population by 4 different groups \citep{jiao2023,zhou2023,wang2023,ou2023}. They did this by dividing space into cells containing many stars. The mean velocity and velocity dispersion of all the stars in each cell is then fed into the Jeans equation to calculate the galactic potential gradient, assuming an axisymmetric potential, and hence the corresponding circular velocity at the location of that cell.

The results are all very similar, and show a steadily falling circular velocity with galactocentric radius \citep{moffat2024}, in contrast to the gas rotation curves, which tend not to decline, and have higher velocities. If the lower stellar velocities are assumed to be indicative of circular motion in the galaxy gravitational field, then the kinematic estimate of the total Milky Way mass is reduced to $2.06\times10^{11}\,M_\odot$ \citep{jiao2023,moffat2024}, and the need to introduce a dark matter component to bind the whole Galaxy is considerably reduced by a factor of ${\sim}\,10$. The reduction may be even greater depending on the mass of a previously undetected hot interstellar gas component \citep{nicastro2016}.

The difference between the Gaia DR3 circular rotation curves and gas rotation curves has been ascribed to unreliability of the Milky Way gas rotation curves due to large random and systematic errors, and uncertainties \citep{zhou2023,jiao2023,moffat2024}. However here we show that in fact the difference is real.

\begin{table*}[t!]
\caption{Gas velocity sources.}
\label{tab:sources}
\centering
\begin{tabular}{llccc}
\hline\hline
Source & Type of source & $N_{\rm data}$ & $R_{\odot,s}$ (kpc) & $\Theta_{0,s}$ (km\,s$^{-1}$) \\
\hline
\citet{blitz1979}           & CO                  & 24  & 10    & 250   \\
\citet{clemens1985}         & CO, HI and HII      & 124 & 8.5   & 220   \\
\citet{fich1989}            & HII                        & 91  & 8.13  & 236.3 \\
\citet{merrifield1992}      & HI, HII                    & 98  & 7.9   & 200   \\
\citet{brand1993}           & HI, HII and CO             & 109 & 8.5   & 220   \\
\citet{turbide1993}         & Young stars                & 5   & 8.5   & 220   \\
\citet{pont1997}            & Classical Cepheids  & 26  & 8.5   & 220   \\
\citet{sofue2009}           & HI                         & 17  & 8.0   & 200   \\
\citet{hou2009}             & HII                        & 49  & 8.5   & 220   \\
\citet{hou2009}             & CO                         & 125 & 8.5   & 220   \\
\citet{honma2012}           & Masers                     & 48  & 8.05  & 238   \\
\citet{bobylev2013}         & SF regions                 & 19  & 8.13  & 236.3 \\
\citet{reid2014}            & Masers         & 14 & 8.34 & 240 \\
\citet{bobylev2015}         & OB stars                   & 82  & 8     & 234   \\
\citet{bobylev2015}         & O stars                    & 70  & 8     & 237   \\
\citet{sun2015}             & CO                         & 72  & 8.13  & 236.3 \\
\citet{reid2019}            & Masers          & 27 & 8.15 & 236 \\
\citet{mroz2019}            & Classical Cepheids         & 150 & 8.09  & 233.6 \\
\citet{hirota2020}          & Masers                     & 189 & 7.92  & 226.7 \\
\citet{zhou2024}            & Classical Cepheids         & 79  & 8.0   & 239   \\
\citet{sofue2025}           & CO, HI                     & 405 & 8.178 & 235.1 \\
\hline
\end{tabular}
\tablefoot{$\Theta_{0,s}$ is the circular velocity of the Local Standard of Rest used in the source, and $R_{\odot,s}$ is the galactocentric radius used for the Solar System. These are required to rescale the data for some of the sources.}
\end{table*}

\begin{table}[ht!]
\caption{Published values of $\Theta_0$ and $R_\odot$.}
\label{tab:theta0}
\centering
\begin{tabular}{lcc}
\hline\hline
Source & $\Theta_0$ (km\,s$^{-1}$) & $R_\odot$ (kpc) \\
\hline
Reid et al. 2009            & 254   & 8.4   \\
Honma et al. 2012           & 238   & 8.05  \\
Reid et al. 2014            & 240   & 8.34  \\
Bobylev et al. 2015         & 234   & 8     \\
Bobylev et al. 2015         & 237   &       \\
Huang et al. 2016           & 240   &       \\
McMillan 2017               & 232.8 & 8.2   \\
Chu et al. 2018             &       & 7.93  \\
Reid et al. 2019            & 236   & 8.15  \\
Eilers et al. 2019          & 229   &       \\
Gravity Collaboration 2019  &       & 8.178 \\
Hirota et al. 2020          & 227   & 7.92  \\
Wang et al. 2021            & 231   &       \\
Zhou, Y. et al. 2023        & 234   &       \\
Gaia Collaboration 2023     & 239   &       \\
\hline
Average values              & 236.3 & 8.13  \\
\hline
\end{tabular}
\end{table}

\section{Averaging of the rotation curves in radial bins}
\label{sec:averaging}

While the Pop~II stellar rotation curves from DR3 are very consistent, gas rotation curves have considerable variability due to streaming motions arising out of spiral shocks \citep{nelson1977,williams2001}, warps \citep{nakanishi2016}, and non-axisymmetries in the potential perturbing the rotational velocity of the gas. This variability, characterised as inaccuracies and unreliability, has been alluded to in order to explain the difference between the Pop~II stellar circular rotation curves and gas rotation curves. However the gas velocity data can be averaged in radial bins using a grand averaging technique \citep{sofue2013} which smooths out the azimuthally varying streaming motions, yielding an estimate of the circular velocity from the mean in each radial bin. Here we averaged the velocity values in radial bins taken from 21 published versions of the gas rotation curve \citep{blitz1979,clemens1985,fich1989,merrifield1992,brand1993,turbide1993,pont1997,sofue2009,hou2009,honma2012,bobylev2013,reid2014,bobylev2015,sun2015,reid2019,mroz2019,hirota2020,zhou2024,sofue2025}.

\subsection{Data sources}
\label{sec:data}

Table~\ref{tab:sources} lists the 21 gas velocity sources. The quantities $\Theta_{0,s}$ and $R_{\odot,s}$ are required to rescale the data for some of the sources, since the azimuthal velocity $V$ of a source is given by
\begin{equation}
V=\frac{R}{R_\odot}\left( \frac{V_r}{\sin l}+\Theta_0 \right),
\label{eq:vaz_outer}
\end{equation}
where $V_r$ is the line-of-sight velocity of the source, $R$ is its galactocentric radius, and $l$ is the galactic longitude. For sources at tangent points inside the solar circle, Eq.~\ref{eq:vaz_outer} reduces to
\begin{equation}
V = V_t + \Theta_0 \sin l\,,
\label{eq:vaz_inner}
\end{equation}
where $V_t$ is the tangent velocity.

The values of $\Theta_0$ and $R_\odot$ used in this paper were 236.3~km\,s$^{-1}$ and 8.13~kpc, which were obtained by averaging the values of these quantities measured by 15 authors in the last 15 years, as listed in Table~\ref{tab:theta0}.

Where the gas velocity source used a different value of $\Theta_0$ from 236.3, or a different value of $R_\odot$ from 8.13, the value of $V$, the gas azimuthal velocity, used in the grand averaging process was rescaled using
\begin{equation}
V = V_s + \Theta_0 \frac{R}{R_{0}} - \Theta_{0,s} \frac{R}{R_{\odot,s}}\,,
\label{eq:rescale}
\end{equation}
where $V_s$ is the quoted value in the source.
The rescaled data and python scripts to perform the Grand average and carry out statistical analysis can be found at DOI.org/10.5281/zenodo.20543192.

\subsection{Grand average of the velocities in radial bins}
\label{sec:grand_avg}

\begin{figure}[ht!]
\centering
\includegraphics[width=\hsize]{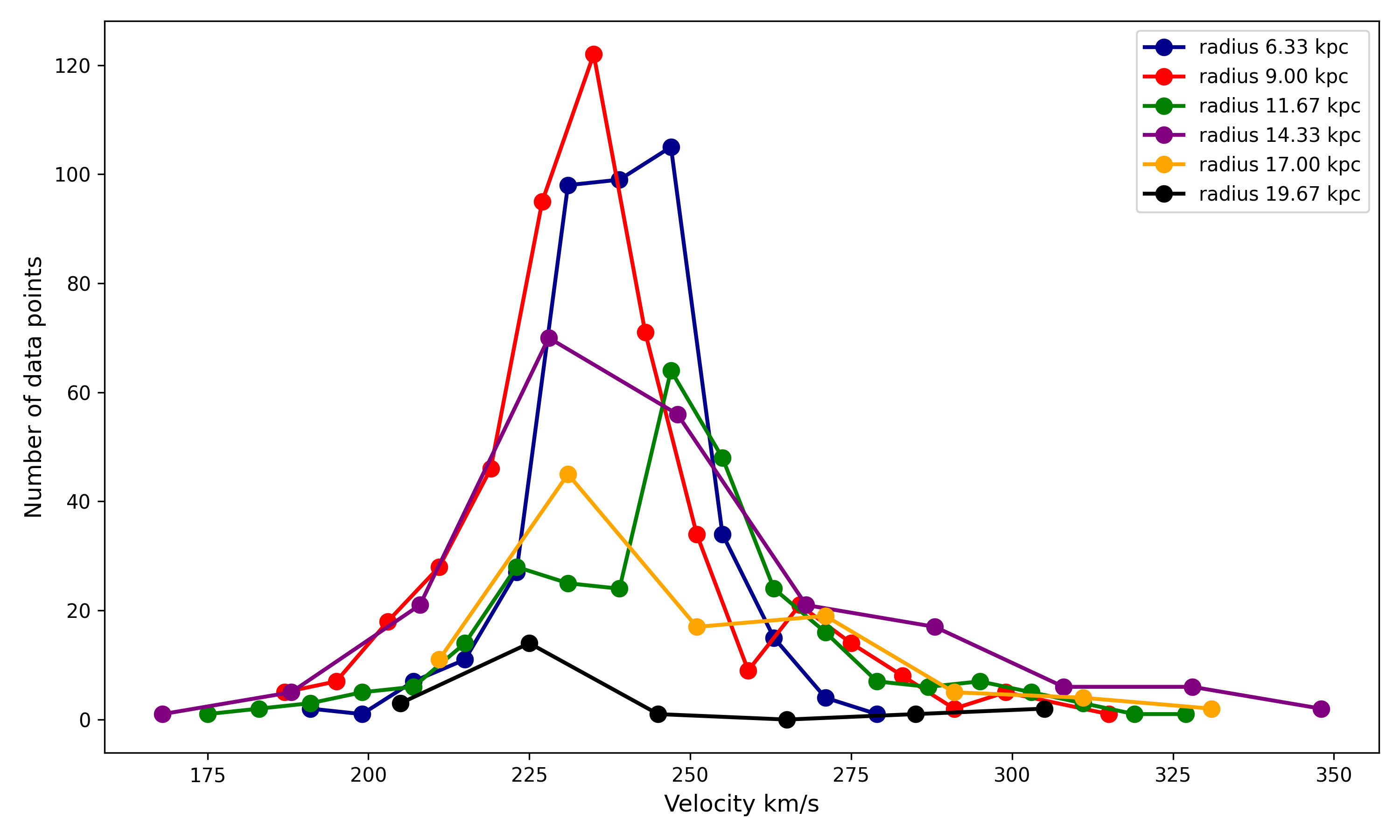}
\caption{Distribution of velocity data points in each of the 6 bins at the indicated radius. The data points are distributed randomly around the bin mean, illustrating that the Grand averaging procedure in the bins identifies the underlying circular velocity of the gas disc at that radius, which is modified by positive or negative streaming perturbations.}
\label{fig:distribution}
\end{figure}

The Grand average is carried out in 6 radial bins with centres from 6.33~kpc to 19.67~kpc, and with bin widths 2.66~kpc. Figure~\ref{fig:distribution} shows the distribution of gas velocity values in each of the 6 radial bins.

\begin{figure}[ht!]
\centering
\includegraphics[width=\hsize]{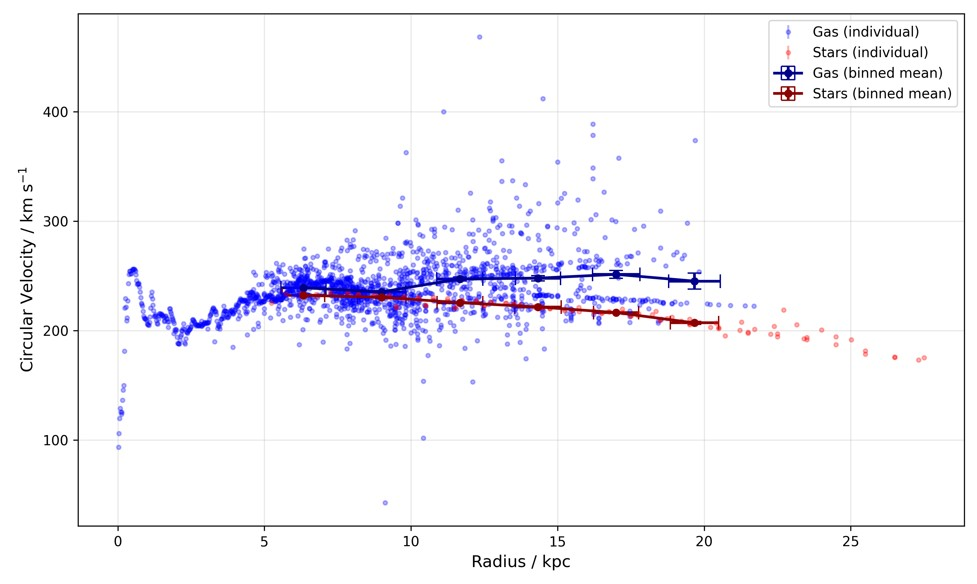}\\
\includegraphics[width=\hsize]{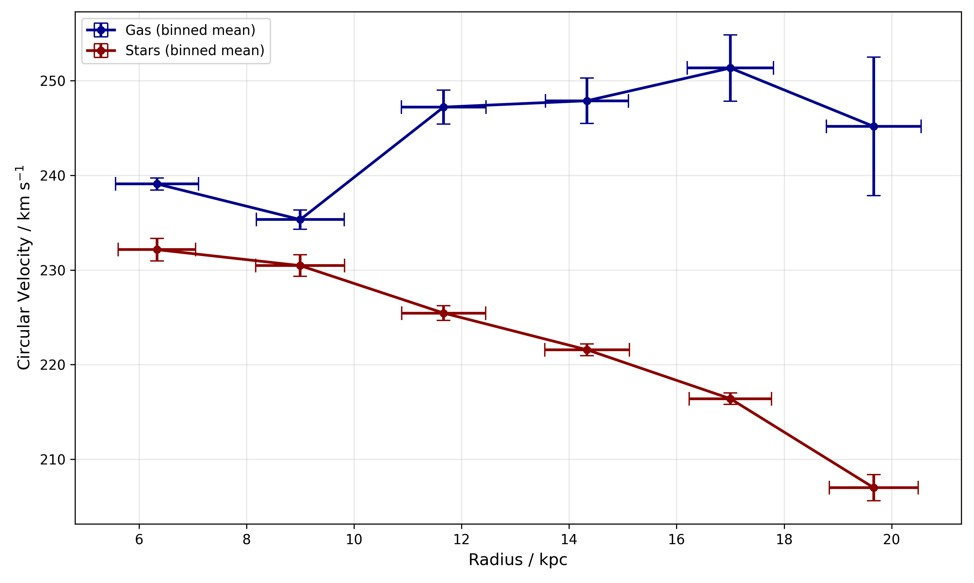}
\caption{Comparison of Gaia Pop~II stars circular velocity rotation curve (red) with the Grand gas rotation curve (blue). Upper panel: full velocity scale on the vertical axis, and individual data points for gas and Pop~II stellar data; lower panel: reduced velocity scale to highlight the size of the gas curve error bars relative to the difference between the rotation curves. Note that while the gas Grand rotation curves are an estimate of the circular velocity (i.e.\ that of a test particle in circular orbit), the individual gas data points are actual gas velocities, while the star data points are the circular velocities from the 4 publications.}
\label{fig:rotation_curves}
\end{figure}

The total number of gas data points in the list of Table~\ref{tab:sources} is $N = 1824$. Each data point has a velocity $V_i$ and a radius $R_i$, $i = 1,\ldots,N$, and these are used to calculate the Grand averaged rotation curve velocity $\bar{V}$ in each bin using the weighted average
\begin{equation}
\bar{V} = \frac{\sum_i w_i V_i}{\sum_i w_i}\,.
\label{eq:vrc}
\end{equation}
For a top-hat average based on radially binned data the weights are
\begin{equation}
w_i = 1 \quad \text{if } R_i \text{ is in the radial bin,}
\end{equation}
or
\begin{equation}
w_i = 0 \quad \text{if } R_i \text{ is not in the radial bin.}
\end{equation}
The standard deviation of the average, representing the spread of values used in the average, is given by
\begin{equation}
\sigma = \sqrt{\frac{\sum_{\rm bin}(V_i - \bar{V})^2}{N_{\rm bin}-1}}\,.
\label{eq:stdev}
\end{equation}
While the standard error of the average is given by
\begin{equation}
\mathrm{stderr} = \frac{\sigma}{\sqrt{N_{\rm bin}}} = \sqrt{\frac{\sum_{\rm bin}(V_i - \bar{V})^2}{N_{\rm bin}(N_{\rm bin}-1)}}\,.
\label{eq:stderr}
\end{equation}
These are the velocity error bars in Fig.~\ref{fig:rotation_curves}, the radial error bars are the standard deviation of the radial positions of the data points about the centre of the bin.

Figure~\ref{fig:rotation_curves} shows the Grand average in the 6 bins of the 21 gas rotation curves plotted alongside the Grand average of the 4 Pop~II Gaia circular velocity rotation curves.

The standard deviation about the mean in each bin is considerably greater (typically ${\sim}\,10\%$ of the azimuthal velocity) than the quoted measurement errors in velocity and radius in the original sources (typically a few per cent), hence these measurement errors are ignored in the averaging process for Figure~\ref{fig:rotation_curves}, though they are used in the averaging for Figure~\ref{fig:rc6abins}.  They are added in when the error is relevant in the statistical analysis comparing the gas and Gaia Pop~II circular velocity curves.

The same Grand averaging is applied separately to the Gaia Pop~II stellar data of which there is a total of 108 data points between the 4 papers.

\section{Statistical significance of the difference}
\label{sec:statistics}

The interstellar gas component and the Pop~II stellar component have distinctly different circular velocity rotation curves. Recall that these are both interpreted to be the circular velocity of a test particle orbiting in the Galaxy's gravitational field in the galactic plane, and should therefore ideally represent the same underlying potential. However, the gas rotation curve sits above the Gaia stellar rotation curve at all radii beyond 5~kpc, and beyond a galactocentric radius of 10~kpc the difference amounts to several standard errors.

\subsection{Welch's t-test on Binned Data}

To assess how distinct the binned rotation curves are statistically, we applied Welch's t-test \citep{bevington2003,nist2012,frayne2015,derrick2016} to the individual radial bins. This test is designed to assess whether two sets of data with different variances nevertheless share the same mean. The velocity data points in each bin are distributed randomly about the bin mean, and hence are amenable to individual Welch's t-tests in each separate bin, since there is no overlap in the radial extents.

The t-statistic at each radial bin for the null hypothesis that the gas and star rotation curves are commensurate is given by
\begin{equation}
t = \frac{\bar{V}_{\rm gas} - \bar{V}_{\rm stars}}{\sqrt{\mathrm{err}_{\rm gas}^2 + \mathrm{err}_{\rm stars}^2}}\,.
\label{eq:tstat}
\end{equation}
Here $\mathrm{err}_{\rm gas}$ and $\mathrm{err}_{\rm stars}$ both have the estimate of the original error in the velocity measurements added in quadrature to the standard error in the bin average.

The degrees of freedom $\nu$ are given by \citep{nist2012,derrick2016}
\begin{equation}
\nu = \frac{(\mathrm{err}_{\rm gas}^2 + \mathrm{err}_{\rm stars}^2)^2}{\mathrm{err}_{\rm gas}^4/(N_{\rm gas}-1) + \mathrm{err}_{\rm stars}^4/(N_{\rm stars}-1)}\,,
\label{eq:dof}
\end{equation}
with the probability of the gas and stars data sets in the bin having the same mean given by \citep{bevington2003}
\begin{equation}
p(t,\nu) = \frac{1}{\sqrt{\nu\pi}}\,\frac{\Gamma[(\nu+1)/2]}{\Gamma[\nu/2]}\,\left(1 + \frac{t^2}{\nu}\right)^{-(\nu+1)/2}\!.
\label{eq:pvalue}
\end{equation}

For individual radial bins we show the $t$-statistic results, and corresponding probability values, in Table~\ref{tab:welch6}.

The most conservative confidence level rejecting the null hypothesis that the circular velocity rotation curves of the gas and Pop~II are commensurate given by this analysis in the bins is 99.61\%. The bins do not overlap, so that the Welch tests in the individual bins are completely independent, and apply at different radii. However if we take the most pessimistic statistical view and apply the Bonferroni adjustment to the probabilities applicable for multiple tests \citep{frayne2015}, i.e.\ multiply the probabilities by the number of tests, 6, then the most conservative confidence level is reduced only to 97.64\%, while the other bins remain at virtually 100\% confidence level.   The null hypothesis that the gas and Pop II stellar tracers share the same mean velocity is rejected with a confidence level > 90\% at all radii.

We show in Appendix~\ref{app:bins} that this conclusion is independent of the number of radial bins used. And we also show that the conclusion also holds if we use weights = $1/orig\_err_{gas}^2$ and $1/orig\_err_{star}^2$ for the bin averages, instead of using equal weights, where $orig\_err_{gas}$ and $orig\_err_{stars}$ are the estimates of the original velocity measurements for gas and stars. 

\begin{table*}[t!]
\caption{Welch's test results in the 6 bins.}
\label{tab:welch6}   
\centering
\begin{tabular}{cccc}
\hline\hline
Radius (kpc) & $t$-statistic & $p$-value & Confidence level (\%) \\
\hline
6.33  & 5.14  & $2.24\times10^{-4}$ & 99.98 \\
9.00  & 3.17  & $3.92\times10^{-3}$ & 99.61 \\
11.67 & 11.09 & $5.98\times10^{-23}$ & ${\sim}\,100$ \\
14.33 & 10.63 & $5.10\times10^{-21}$ & ${\sim}\,100$ \\
17.00 & 9.86  & $1.89\times10^{-16}$ & ${\sim}\,100$ \\
19.67 & 5.13  & $2.67\times10^{-5}$ & 99.99 \\
\hline
\end{tabular}
\end{table*}

\subsection{Error-in-Variables Regression and Physical Divergence}

To verify the robustness of the statistical divergence without relying on spatial binning, and to characterize the radial variation of the discrepancy, we applied an Error-in-Variables (EIV) regression to the unbinned datasets \citep{fuller1987, tremaine2002, kelly2007}. By fitting a continuous cubic spline to the Gaia stellar kinematics, we calculated the velocity residual $\Delta V_i = V_{{\rm gas},i} - V_{\rm spline}(R_i)$ at the exact galactocentric radius $R_i$ of each gas measurement. We then modeled this dataset using a linear radial gradient, $\Delta V = \alpha R + \beta$.

The EIV regression incorporates both velocity measurement errors and radial distance uncertainties. The full details of the procedure are provided in Appendix~\ref{app:eiv}. We found a positive radial gradient of $\alpha = 2.54 \pm 0.19 \text{ km s}^{-1}\text{kpc}^{-1}$, and an offset of $\beta = -13.27 \pm 2.02 \text{ km s}^{-1}$, across the full radial domain where the gas and stellar datasets overlap, 5.24 to 21.7 kpc. A statistical comparison between a constant offset model (with $\alpha$ set to zero) and the linear gradient model yields a $p$-value of $<0.0001$. This gives a high degree of confidence to the conclusion that this trend in $\Delta V$ is not a constant calibration offset, but increases significantly with galactocentric radius.

Furthermore, the EIV analysis yields an intrinsic scatter in $\Delta V$ of $22.98 \text{ km s}^{-1}$. If the divergence between the datasets were driven purely by random instrumental measurement noise, the intrinsic scatter and the radial gradient ($\alpha$) would approach zero. Their substantial values confirm that the discrepancy is a radially dependent, genuine physical phenomenon.

\section{Conclusion --- Implications for the Milky Way and external galaxies}
\label{sec:conclusion}

The Milky Way has only one gravitational field determined by whatever mass it contains. If the gas rotation curve represented the circular velocity in the Milky Way gravitational field, then how could the stellar disc be in a steady state in that field? With too low a velocity the stellar disc would collapse towards the centre of the galaxy. The most likely explanation for the distinct rotation curves is that the rotation curve of Pop~II stars represents the steady circular velocity in that gravitational field, and the higher gas rotation curve has an alternative explanation.

This would imply that the Galaxy mass from the stellar rotation curve is considerably less than previous estimates based on the gas rotation curve, for instance \citet{jiao2023} in Fig.~8 estimated the mass at 26.5~kpc to be ${\sim}\,2\times10^{11}\,M_\odot$. From Fig.~5 of \citet{jiao2023} the fraction of this which is baryonic matter is 29\%, and the fraction which is deemed to be dark matter is 71\%, with the mass of each component being proportional to the square of the associated velocity. So the ratio of dark mass to baryonic mass is 2.5. However, if the Grand gas rotation curve velocity was taken to be the circular velocity at 26.5~kpc, ${\sim}\,245$~km\,s$^{-1}$ from Fig.~\ref{fig:rotation_curves} assuming a flat rotation curve beyond 20~kpc, then the total mass at that radius would increase to $3.7\times10^{11}\,M_\odot$, with the baryonic fraction 16\% (assuming the same baryonic mass). Hence the dark matter fraction (given by the total mass minus the baryonic mass) increases to 84\%. The ratio of dark mass to baryonic mass would then be 5.4, which implies that using the Gaia DR3 rotation curve the fraction of dark matter decreases by a factor of 2.2.

On the other hand at the radius of 15~kpc the mass estimate from \citet{jiao2023} is ${\sim}\,1.7\times10^{11}\,M_\odot$, with 34\% in baryons, and 66\% in dark matter. So the ratio of dark mass to baryonic mass at 15~kpc is 1.9. But using ${\sim}\,248$~km\,s$^{-1}$ from the gas curve in Fig.~\ref{fig:rotation_curves}, the total mass at that radius would be $2.2\times10^{11}\,M_\odot$, and the baryon and dark fractions would be 27\% and 73\% respectively. In that case the ratio of dark mass to baryonic mass at that radius would be 2.7, and the decrease in the dark matter fraction using the Gaia rotation curve at 15~kpc would be by a factor of 1.4.

The decrease in the dark matter fraction implied by the Gaia results consequently increases with radius, from 1.4 at 15~kpc to 2.2 at 26.5~kpc. This is consistent with the claimed decrease by a factor of 10 in the dark matter mass for the whole Galaxy \citep{jiao2023,moffat2024}.

Consequently, if this interpretation of the Gaia rotation curves is correct, i.e. that they represent the circular velocity in the Galaxy, then the dark matter masses of external galaxies based on gas rotation curves are a considerable overestimate.

\section{Discussion}
\label{sec:discussion}

Alternative explanations for the higher gas velocities are unclear. The gas disc is corrugated inside the solar circle, and warped outside a galactocentric radius of $\sim 10$ kpc \citep{nakanishi2016}, but the grand averaging procedure is designed to deal with any consequent deviations from circular motion. There may however be secular changes to rotation, particularly from shocks in spiral arms, where energy and angular momentum are lost from the gas flow. However numerical simulations of the formation of galactic discs to-date have not demonstrated such a secular effect \citep[][Fig.~8]{williams2001}.

An alternative explanation of the higher velocity of the gas rotation curve could be the effect of the large scale magnetic field of the Galaxy. This could either speed up the gas disc \citep{nelson1988}, which would require the gas to move radially out of the Galaxy, though in an outer region where the density of the gas is relatively low. Or the field could provide an inward force, as in a Z-Pinch \citep{battaner1992,kutschera2004,ruizgranados2010,ruizgranados2012}, supplementing the Galactic gravitational force. All disc galaxies are known to have magnetic fields, typically with an energy density comparable to the thermal and turbulent energy density of the interstellar gas, so a similar magnetic effect may be present in external galaxies.

\bibliographystyle{aa}
\bibliography{references}

\begin{appendix}

\section{Independence of the conclusion on the number of radial bins, and on the weights in the bin averages}
\label{app:bins}

To check that the conclusion regarding the significance of the result is not dependent on the number of radial bins used, we carried out the Grand averaging and statistical analysis with respectively 8 and 10 bins, with the following results.

\begin{figure}[ht!]
\centering
\includegraphics[width=\hsize]{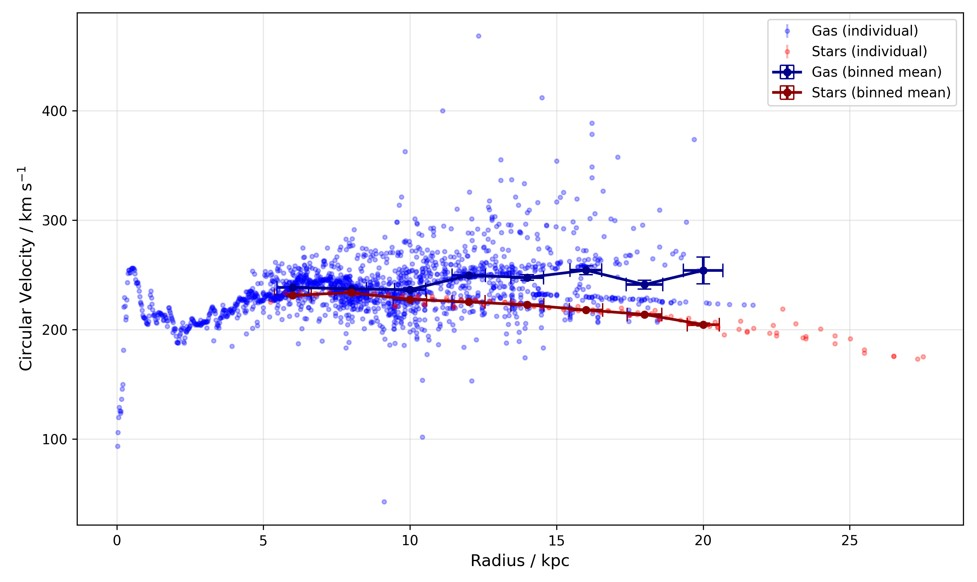}\\
\includegraphics[width=\hsize]{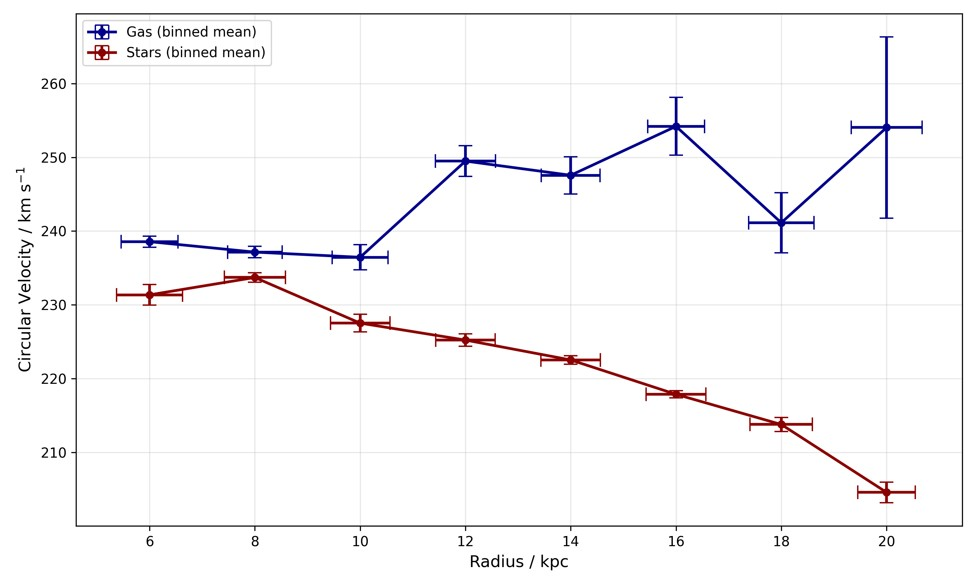}
\caption{Same as Fig.~\ref{fig:rotation_curves}, but for 8 bins.}
\label{fig:rc8bins}
\end{figure}

For 8 radial bins of width 2~kpc we show the $t$-statistic results, and corresponding probability values, in Table~\ref{tab:welch8}, and the rotation curves in Fig.~\ref{fig:rc8bins}.

\begin{table}
\caption{Welch's test results in the 8 bins.}
\label{tab:welch8}
\centering
\begin{tabular}{cccc}
\hline\hline
Radius (kpc) & $t$-statistic & $p$-value & Confidence level (\%) \\
\hline
6    & 4.57  & $1.13\times10^{-3}$ & 99.88 \\
8    & 3.41  & $2.09\times10^{-3}$ & 99.79 \\
10   & 4.31  & $7.85\times10^{-5}$ & 99.99 \\
12   & 10.77 & $6.07\times10^{-21}$ & ${\sim}\,100$ \\
14   & 9.68  & $1.21\times10^{-17}$ & ${\sim}\,100$ \\
16   & 9.20  & $1.19\times10^{-14}$ & ${\sim}\,100$ \\
18   & 6.55  & $3.48\times10^{-8}$ & ${\sim}\,100$ \\
20   & 4.00  & $1.52\times10^{-3}$ & 99.84 \\
\hline
\end{tabular}
\end{table}

\begin{figure}[ht!]
\centering
\includegraphics[width=\hsize]{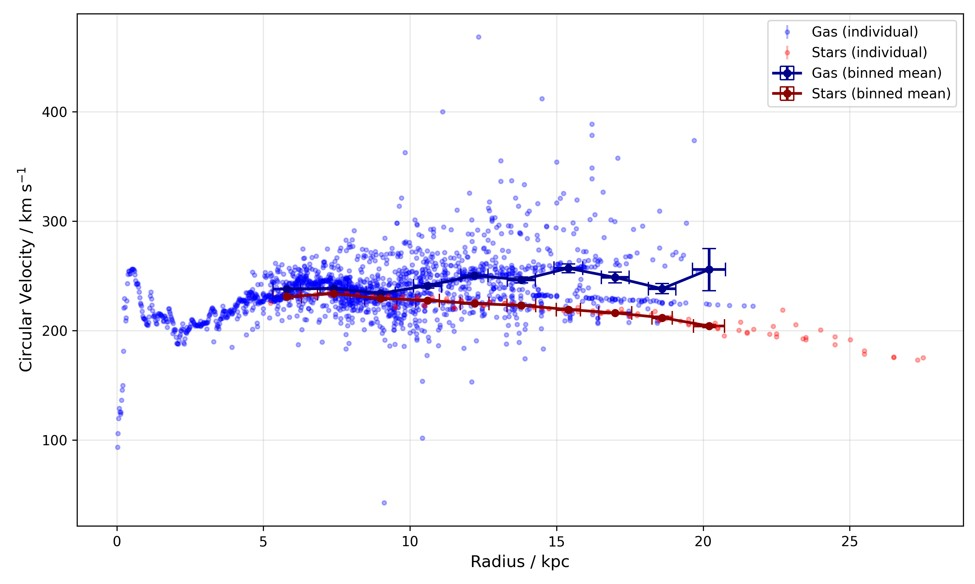}\\
\includegraphics[width=\hsize]{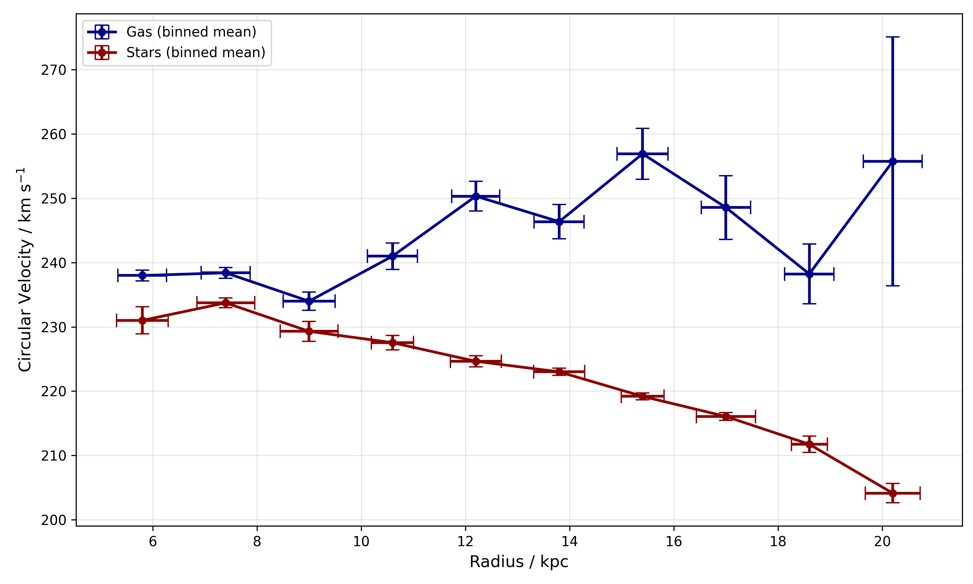}
\caption{Same as Fig.~\ref{fig:rotation_curves}, but for 10 bins.}
\label{fig:rc10bins}
\end{figure}

For 10 radial bins of width 1.6~kpc we show the $t$-statistic results, and corresponding probability values, in Table~\ref{tab:welch10}, and the rotation curves in Fig.~\ref{fig:rc10bins}.

\begin{table}
\caption{Welch's test results in the 10 bins.}
\label{tab:welch10}
\centering
\begin{tabular}{cccc}
\hline\hline
Radius (kpc) & $t$-statistic & $p$-value & Confidence level (\%) \\
\hline
5.8  & 3.09  & $1.80\times10^{-2}$ & 98.21 \\
7.4  & 4.11  & $3.95\times10^{-4}$ & 99.96 \\
9.0  & 2.22  & $3.76\times10^{-2}$ & 96.24 \\
10.6 & 5.72  & $2.50\times10^{-7}$ & ${\sim}\,100$ \\
12.2 & 10.33 & $4.79\times10^{-19}$ & ${\sim}\,100$ \\
13.8 & 8.57  & $2.00\times10^{-14}$ & ${\sim}\,100$ \\
15.4 & 9.41  & $1.90\times10^{-14}$ & ${\sim}\,100$ \\
17.0 & 6.51  & $3.80\times10^{-8}$ & ${\sim}\,100$ \\
18.6 & 5.51  & $5.20\times10^{-6}$ & ${\sim}\,100$ \\
20.2 & 2.66  & $2.36\times10^{-2}$ & 97.64 \\
\hline
\end{tabular}
\end{table}

For these  the binned averages in Figures 2, A1, and A2 equal weights were given to all the gas and stellar data points.  A standard procedure is often to use the inverse squared measurement errors in the velocities as weights in the averages, though when the errors are small compared to the velocities this can give undue bias to the data with the smallest errors, distorting the result.   Nevertheless we carried out the analysis again with 6 bins using the inverse squared errors as weights, i.e. using in equation (4) to obtain the velocity averages with weights

\begin{equation}
w_i = 1 /orig\_err^2 \quad \text{if } R_i \text{ is in the radial bin,}
\end{equation}
where $orig\_err$ is the estimate of the measurement error quoted in the sources,\\

or
\begin{equation}
w_i = 0 \quad \text{if } R_i \text{ is not in the radial bin.}
\end{equation}

\begin{figure}[ht!]
\centering
\includegraphics[width=\hsize]{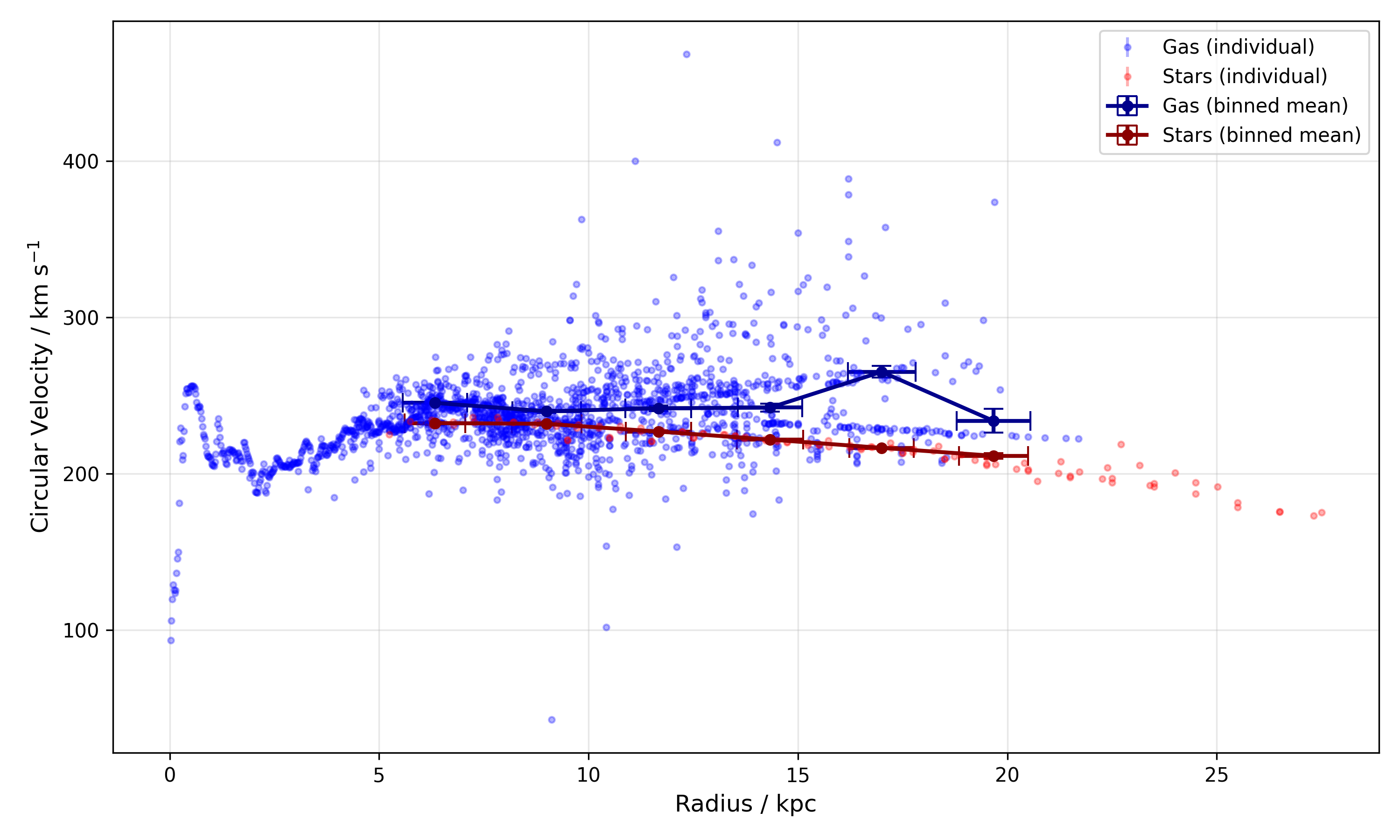}\\
\includegraphics[width=\hsize]{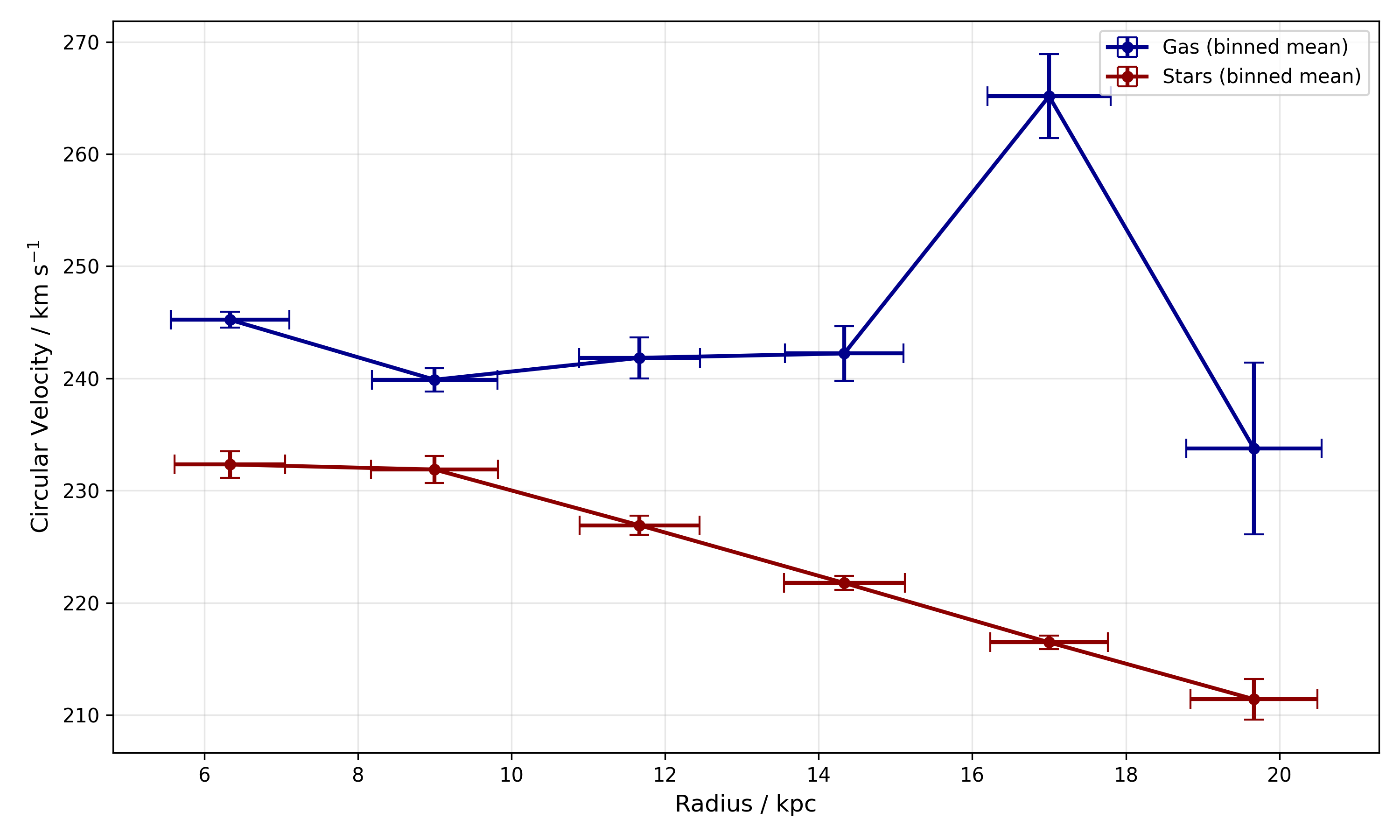}
\caption{Same as Fig.~\ref{fig:rotation_curves}, but with A.1 and A.2 weights.}
\label{fig:rc6abins}
\end{figure}

Equations (7) and (8) then yield $\sigma$ and stderr, to use in the Welch test, that is equations (9), (10) , and (11).
For 6 radial bins of width 2.66~kpc with inverse squared error weights we show the $t$-statistic results, and corresponding probability values, in Table~\ref{tab:welch6a}, and the rotation curves in Fig.~\ref{fig:rc6abins}.
\begin{table}
\caption{Welch's test results in 6 bins with inverse squared error weights.}
\label{tab:welch6a}
\centering
\begin{tabular}{cccc}
\hline\hline
Radius (kpc) & $t$-statistic & $p$-value & Confidence level (\%) \\
\hline
6.33  & 9.30  & $2.99\times10^{-7}$ & 99.99 \\
9.00  & 5.00  & $1.92\times10^{-5}$ & 99.99 \\
11.67 & 7.38 & $9.20\times10^{-12}$ & ${\sim}\,100$ \\
14.33 & 8.16 & $9.20\times10^{-14}$ & ${\sim}\,100$ \\
17.00 & 12.84  & $2.79\times10^{-23}$ & ${\sim}\,100$ \\
19.67 & 2.85  & $9.85\times10^{-3}$ & 99.01 \\
\hline
\end{tabular}
\end{table}

\section{Error-in-Variables Regression Procedure}
\label{app:eiv}

To rigorously quantify the divergence between the gas and stellar rotation curves without relying on spatial binning, we employed an Error-in-Variables (EIV) regression \citep{fuller1987, tremaine2002, kelly2007}. This method allowed us to evaluate the velocity difference $\Delta V$ at the exact location of every gas data point while properly propagating uncertainties in both velocity and galactocentric radius.

First, we established a continuous baseline for the stellar kinematics by fitting a cubic spline $V_{\rm spline}(R)$ to the Gaia DR3 stellar rotation data. For each individual gas measurement $i$, characterized by radius $R_i$ and velocity $V_{{\rm gas},i}$, we computed the discrepancy
\begin{equation}
\Delta V_i = V_{{\rm gas},i} - V_{\rm spline}(R_i).
\end{equation}

Because the stellar rotation curve exhibits a radial gradient, any uncertainty in the measured distance to a gas cloud $\sigma_{R,{\rm gas},i}$ induces a corresponding uncertainty in the expected stellar velocity at that location. Following the Maximum Likelihood framework for linear regression with intrinsic scatter outlined by \citet{tremaine2002}, the uncertainty in the radial coordinate propagates into the residual variance as a function of the modeled slope, $\alpha$. The total variance for each data point is therefore
\begin{equation}
\sigma_{{\rm total},i}^2 = \sigma_{V,{\rm gas},i}^2 + \sigma_{V,{\rm spline},i}^2 + \left[\left.\frac{dV_{\rm spline}}{dR}\right|_{R_i} + \alpha \right]^2 \sigma_{R,{\rm gas},i}^2 + \sigma_{\rm int}^2.
\end{equation}
Here, $\sigma_{\rm int}$ represents the intrinsic velocity dispersion of the gas not arising from measurement or interpolation errors. 

To evaluate this total variance, we established a representative error budget. For the observational gas velocity errors $\sigma_{V,{\rm gas},i}$ (= $orig\_err_{gas}$)we utilized the published measurement uncertainties from the original literature. Because individual distance uncertainties are not uniformly tabulated across all historical datasets, we adopted a standard, representative radial uncertainty of $\sigma_{R,{\rm gas},i} = 0.05 R_i$ (a $5\%$ distance error), which is consistent with typical modern parallax and kinematic distance limits. Finally, we assigned a nominal uncertainty of $\sigma_{V,{\rm spline},i} = 1.0 \text{ km s}^{-1}$ to account for minor interpolation variances between the discrete Gaia stellar data points.

We determined the model parameters for a linear gradient ($\Delta V(R) = \alpha R + \beta$) and the intrinsic scatter $\sigma_{\rm int}$ simultaneously by forcing the reduced chi-squared statistic of the residuals to unity. This was achieved by minimizing the objective function:
\begin{equation}
\left(\chi_{\rm red}^2 - 1\right)^2 = \left[ \frac{1}{N-2} \sum_{i=1}^{N} \frac{\left[\Delta V_i - (\alpha R_i + \beta)\right]^2}{\sigma_{{\rm total},i}^2} - 1 \right]^2
\end{equation}
where $N$ is the total number of gas data points. The statistical preference for the radially dependent linear model over a constant offset model ($\alpha = 0$) was confirmed by evaluating the $p$-value of the optimal gradient $\alpha$. This was computed using a two-tailed $t$-test against the null hypothesis that the true physical slope is zero, demonstrating with high confidence that the separation between the gas and stellar curves is a radially increasing physical feature rather than a uniform bias.
 
\end{appendix}

\end{document}